\documentclass[twocolumn,longauthor]{aastex63}

\newcommand{\gaia}{{\it Gaia} }

\usepackage{booktabs, tabularx, makecell, multirow, xspace}
\newcolumntype{C}[1]{>{\centering\let\newline\\\arraybackslash\hspace{0pt}}m{#1}}

\definecolor{colorTC}{rgb}{.2,.7,.2}

\shorttitle{Machine Learning the $6^{\text{th}}$ Dimension}
\shortauthors{Dropulic et al.}

\graphicspath{{./}{figures/}}
\usepackage{url}
\usepackage[normalem]{ulem}
\usepackage{amsmath}
\usepackage{comment}
\usepackage{svg}
\usepackage{natbib}

\begin{document}

\title{\vspace{-40pt} Machine Learning the $\mathbf{6^{\text{th}}}$ Dimension: \\Stellar Radial Velocities from 5D Phase-Space Correlations}

\correspondingauthor{Adriana Dropulic}
\email{dropulic@princeton.edu}

\author[0000-0002-7352-6252]{Adriana Dropulic}
\affiliation{Department of Physics, Princeton University, Princeton, NJ 08544, USA }

\author[0000-0002-0376-6461]{Bryan Ostdiek}
\affiliation{Department of Physics, Harvard University, Cambridge, MA 02138, USA}
\affiliation{The NSF AI Institute for Artificial Intelligence and Fundamental Interactions}

\author[0000-0001-8590-2043]{Laura~J. Chang}
\affiliation{Department of Physics, Princeton University, Princeton, NJ 08544, USA }

\author[0000-0003-2486-0681]{Hongwan Liu}
\affiliation{Department of Physics, Princeton University, Princeton, NJ 08544, USA }
\affiliation{Center for Cosmology \& Particle Physics, Department of Physics, New York University, New York, NY 10003, USA}

\author[0000-0002-7040-3038]{Timothy Cohen}
\affiliation{Institute for Fundamental Science, Department of Physics, University of Oregon, Eugene, OR 97403, USA }

\author[0000-0002-8495-8659]{Mariangela Lisanti}
\affiliation{Department of Physics, Princeton University, Princeton, NJ 08544, USA }

\begin{abstract}
The \gaia satellite will observe the positions and velocities of over a billion Milky Way stars.  In the early data releases, the majority of observed stars do not have complete 6D phase-space information.  In this Letter, we demonstrate the ability to infer the missing line-of-sight velocities until more spectroscopic observations become available.  We utilize a novel neural network architecture that, after being trained on a subset of data with complete phase-space information, takes in a star's 5D astrometry (angular coordinates, proper motions, and parallax) and outputs a predicted line-of-sight velocity with an associated uncertainty.  Working with a mock \gaia catalog, we show that the network can successfully recover the distributions and correlations of each velocity component for stars that fall within $\sim 5$ kpc of the Sun.  We also demonstrate that the network can accurately reconstruct the velocity distribution of a kinematic substructure in the stellar halo that is spatially uniform, even when it comprises a small fraction of the total star count. 
\vspace{30pt}
\end{abstract}

\section{Introduction}\label{sec:intro} 

\gaia has ushered in a new age in astrometry, with the goal of providing precise positions and velocities for an unprecedented number of stars in the Milky Way~\citep{2016A&A...595A...1G,gaia_collaboration_gaia_2018}.  This complete phase-space information will revolutionize our understanding of both disk and halo dynamics.  In the current data release~(EDR3), most \gaia stars only have 5D astrometry available (two angular coordinates, two proper motions, and parallax); less than 1\% of the stars  have a measured line-of-sight velocity~\citep{gaiacollaboration2020gaia}. 
We demonstrate how to use regressive neural networks to successfully predict stellar line-of-sight velocities with an associated uncertainty from 5D astrometry.  This approach increases the  scientific output of early \gaia data releases until more spectroscopic data is available. 

The science applications that benefit from having line-of-sight velocities are vast---see \cite{2005MNRAS.359.1306W} for a review---and include measuring the Milky Way potential, obtaining the local dark matter density, distinguishing the thin and thick disk, and mapping substructure in the stellar disk.  One particular case where having full stellar phase-space information is highly beneficial is the identification of stellar remnants of disrupted satellite galaxies in the Milky Way~\citep{1996ApJ...465..278J,1998ApJ...495..297J,Bullock_2005}.  Such mergers are a natural consequence of  hierarchical structure formation~\citep{1978MNRAS.183..341W}, and in addition to depositing stars that form the stellar halo, also leave behind dark matter substructures.

The neural network approach proposed in this Letter has the potential to dramatically and immediately increase the subset of \gaia data that can be used towards these applications. 
We present the results of training, validating, and testing the network using a simulated \gaia mock catalog that models both the smooth stellar halo and disk, as well as kinematic substructure in the halo.  The network is trained on a subset of stars with full 6D phase-space information using a simple network loss function similar to the method described in~\cite{ieee374138}. While 5D astrometry alone is insufficient to provide a reliable prediction of the line-of-sight velocity for every individual star, incorporating the learned uncertainty allows us to obtain accurate \textit{distributions} of the line-of-sight velocity for the full population of stars, as well as the correlations between different velocity components in the Galactocentric frame. Additionally, the confidence of the network prediction of the line-of-sight velocity of each star can be inferred from the learned uncertainty.

 The machine learning approach introduced in this Letter is intended to work in tandem with spectroscopic surveys, such as
APOGEE, the \emph{Gaia}-ESO-Survey, GALAH, LAMOST, RAVE, and SEGUE~\citep{2017AJ....154...94M, 2012Msngr.147...25G, 2020arXiv201102505B,2012RAA....12.1197C, kunder_radial_2017, 2009AJ....137.4377Y}.  These surveys provide high-quality line-of-sight velocities, as well as abundances and stellar parameters.  However, their sky coverage typically  overlaps with only a small fraction of the \gaia catalog.  While direct observations remain the gold standard, we show that neural network-based inference can serve a complementary role in studying the Milky Way's stellar phase-space distribution.

This paper is organized as follows. Section \ref{sec:methods} introduces the simulated mock \gaia catalog, and overviews the machine learning architecture and the training procedure. Section \ref{sec:results} summarizes the network's success in predicting stellar line-of-sight velocities.  We conclude in Section~\ref{sec:conlude} and include three appendices that are referenced in the text. The neural network used in this work is provided at the following \texttt{github} repository \url{https://github.com/adropulic/ML\_6th\_Dimension}.

\section{Methodology}\label{sec:methods}
In this section, we provide a description of the mock dataset used in this work, followed by a detailed discussion of the neural network setup and training strategy.
\subsection{Mock Data Catalog}\label{sec:simdata}
To build our mock catalog, we start from the mock \gaia~DR2 catalog introduced in~\cite{Rybizki_2018}. This catalog was generated by applying a 3D dust extinction map to mock Milky Way stellar data simulated using the public code \textsc{Galaxia}~\citep{sharma_galaxia_2011}. \textsc{Galaxia} employs the Besan\c{c}on Galactic model~\citep{2003A&A...409..523R}, which includes the bulge, thin and thick disk, as well as the stellar halo.  The  final mock catalog of $\sim1.6$ billion stars, each with full astrometric and photometric properties, was created by applying the combined 3D extinction map from \cite{Bovy_2016} to model dust attenuation along with the \gaia selection criteria.  Mock stars were populated by directly sampling the analytic phase-space distribution. This final point is important for our application, as the network should ideally learn the original kinematic distribution, not artifacts that may have been introduced in the sampling procedure that populates the mock catalog. Comparatively, generating mock catalogs from numerical galaxy simulations with finite stellar-mass resolution can introduce artifacts into the stellar kinematics.  

For this study, we begin with the full \textsc{Galaxia} mock catalog of stars\footnote{\url{https://dc.zah.uni-heidelberg.de/gdr2mock/q/download/static/}}  and select only those with relative parallax uncertainty $\delta\varpi/\varpi<0.1$ and line-of-sight velocity $v_{\text{los}} \in [-550, 550]$~km/s.  This results in a sample of $\sim75$ million stars, concentrated within $\sim 5$~kpc of the Sun.  This sample only models the smooth component of the stellar halo and disk.  To study the effect of kinematic substructure on the neural network learning, we supplement the \textsc{Galaxia} catalog with a population of spatially uniform  stars whose velocity distribution resembles that of \gaia Enceladus~\citep{2018MNRAS.478..611B,2018Natur.563...85H}.  The Enceladus-like stars are introduced into the training, validation, and test sets by replacing 50\% of stars with low-metallicity ([Fe/H] $< -1.3$) in each of these sets with stars drawn from the Enceladus distribution in~\cite{Necib:2018iwb}.

The training and validation sets contain seven and one million randomly selected stars, respectively, both with $G < 13.5$ and $T_\text{eff} \in [3550, 6900]$~K. This is intended to parallel the subset of \textit{Gaia} data with full 6D information~\citep{katz_gaia_2019}. The test set consists of $\sim 10$ million stars of any magnitude and temperature.
 
Approximately $0.5\%$ of the training and validation set stars and $0.3\%$ of the test set stars are in substructure. Only $\sim 14\%$ of stars in the test set fall into the magnitude and effective temperature range of the training set, including the substructure stars.  

\subsection{Neural Network Architecture \& Training}\label{sec:network}

A feedforward neural network is trained to predict a line-of-sight velocity for each star in the input catalog, with an associated uncertainty on this prediction. The network is implemented in \texttt{Keras}~\citep{chollet2015keras} using the \texttt{Tensorflow} backend~\citep{abadi_tensorflow_2016}. It is a combination of two halves that are structured identically except for the last layer (Fig.~\ref{fig:networks}). This compound network structure permits two outputs: the line-of-sight velocity as well as the uncertainty on the network's prediction. 
Each half of the network consists of six layers: the input, four hidden layers, and the output. The input consists of five quantities per star: Galactic longitude ($\ell$), Galactic latitude ($b$), proper motion in right ascension ($\mu_\alpha$), proper motion in declination ($\mu_\delta$), and parallax ($\varpi$). The hidden layers comprise 30 nodes each, and use a hyperbolic tangent activation function. All of the network weights are initialized with the \texttt{glorot uniform} method~\citep{pmlr-v9-glorot10a}. Comprehensive comparisons of different network architectures were performed to determine this optimal setup (Appendix \ref{sec:appendix_network}). For example, we trained the network on only Galactic positions $x, y, z$, and found suboptimal results.

The output layer from one half of the network, which we will call the ``velocity predictor," consists of a single node with linear activation in order to attain a continuous value, the line-of-sight velocity. The output layer from the other half of the network, which we will call the ``uncertainty predictor," uses a ReLU activation function in order to constrain the uncertainty on the predicted line-of-sight velocity to positive values. This network is similar to the mixture density network presented in~\cite{bishopmixturedensity}.

The goal of training is to minimize the weighted Gaussian log-likelihood loss function:
\begin{equation}
    \mathcal{L} =\sum_{i=1}^{N} \frac{w_{i}}{N} \left [ \frac{\left( v_{\text{los},i} - v_{\text{los},i}^{\text{pred}} \right)^2}{\left( \sqrt{2} \sigma_{\text{los}, i}^\text{pred} \right)^2} - \text{log}\left(\frac{1}{\sqrt{2 \pi } \, \sigma_{\text{los,} i}^{\text{pred}} }\right)\right] \,,
    \label{eq:lossfunc}
\end{equation}
where $v_{\text{los}, i}$ is the true value of a star's line-of-sight velocity, $v_{\text{los},i}^\text{pred}$ is the network's predicted value of a star's line-of-sight velocity, $w_{i}$ is the sample weight of a star, $\sigma_{\text{los}, i}^\text{pred}$ is the network's predicted uncertainty, and $N$ is the total number of stars. The sample weights are a function of both $\ell_i$ and $v_{\text{los}, i}$, and are used to force the network to learn the tails of the velocity distribution (Appendix \ref{sec:appendix_network}).

The velocity predictor is first trained using $\sigma_{\text{los}, i}^\text{pred} = 1$ for the loss in Eq.~\eqref{eq:lossfunc}. Then the uncertainty predictor is trained using the $v_{\text{los}, i}$ prediction from the velocity predictor in  Eq.~\eqref{eq:lossfunc} without allowing the velocity predictor to update.
Finally, both the velocity predictor and the uncertainty predictor are updated simultaneously using both the predicted $v_{\text{los},i}^\text{pred}$ and $\sigma_{\text{los}, i}^\text{pred}$ in the loss function. This produces a predicted line-of-sight velocity value and an uncertainty value per star.

\begin{figure*}[t]
\centering
\includegraphics[width=\textwidth]{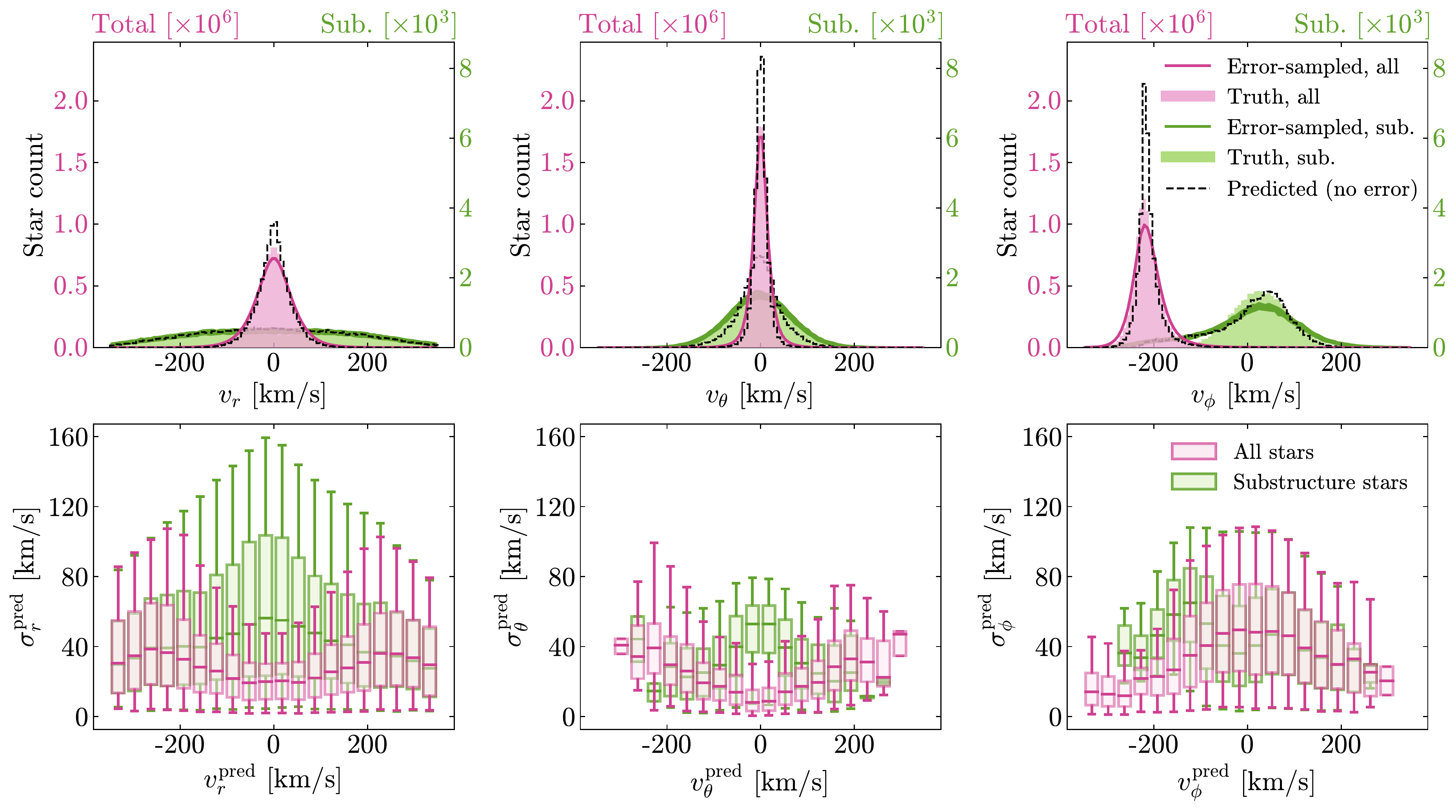}
\caption{The top row shows the velocity distribution for Galactocentric velocity components. The solid pink(green) histogram is the true distribution of all(substructure) test-set stars sampled from the mock catalog; note the separate scales for the separate star counts. The network is trained on 7~million stars($\sim$ 35,000 substructure stars) from the same catalog. 
The black dashed lines show the network's predicted distribution. The dark pink(green) solid line shows the network prediction after it is Gaussian-sampled with the predicted uncertainty. The thickness of this line refers to the minimum and maximum of 50 Monte Carlo trials, described in the text. The error-sampled distribution provides a better approximation of the truth than the predicted distribution alone.  In the bottom row, the predicted uncertainty on the Galactocentric velocities is shown as a function of the predicted velocity. The pink(green) boxes mark the 50\% containment about the median, and the whiskers mark the 5\% and 95\% containment for all(substructure) stars.}
\label{fig:1dhist}
\end{figure*}

During a training epoch, the stars are partitioned into batches of $10^4$ stars, and the network is optimized on a given batch.  The \texttt{Adam} optimizer \citep{kingma_adam_2017} is used to minimize the loss function in Eq.~(\ref{eq:lossfunc}).  

We use an initial learning rate of $10^{-3}$ while the default values are used for the other parameters of the optimizer. If the loss computed on the validation stars does not improve for 10 epochs, the learning rate is decreased by a factor of 10, with a minimum learning rate of $10^{-5}$. Training is stopped when the validation loss has not improved for 40 epochs. We verified that the training and validation losses were similar to check for overfitting.
 
\begin{figure*}[!t]
\centering
\includegraphics[width=\textwidth]{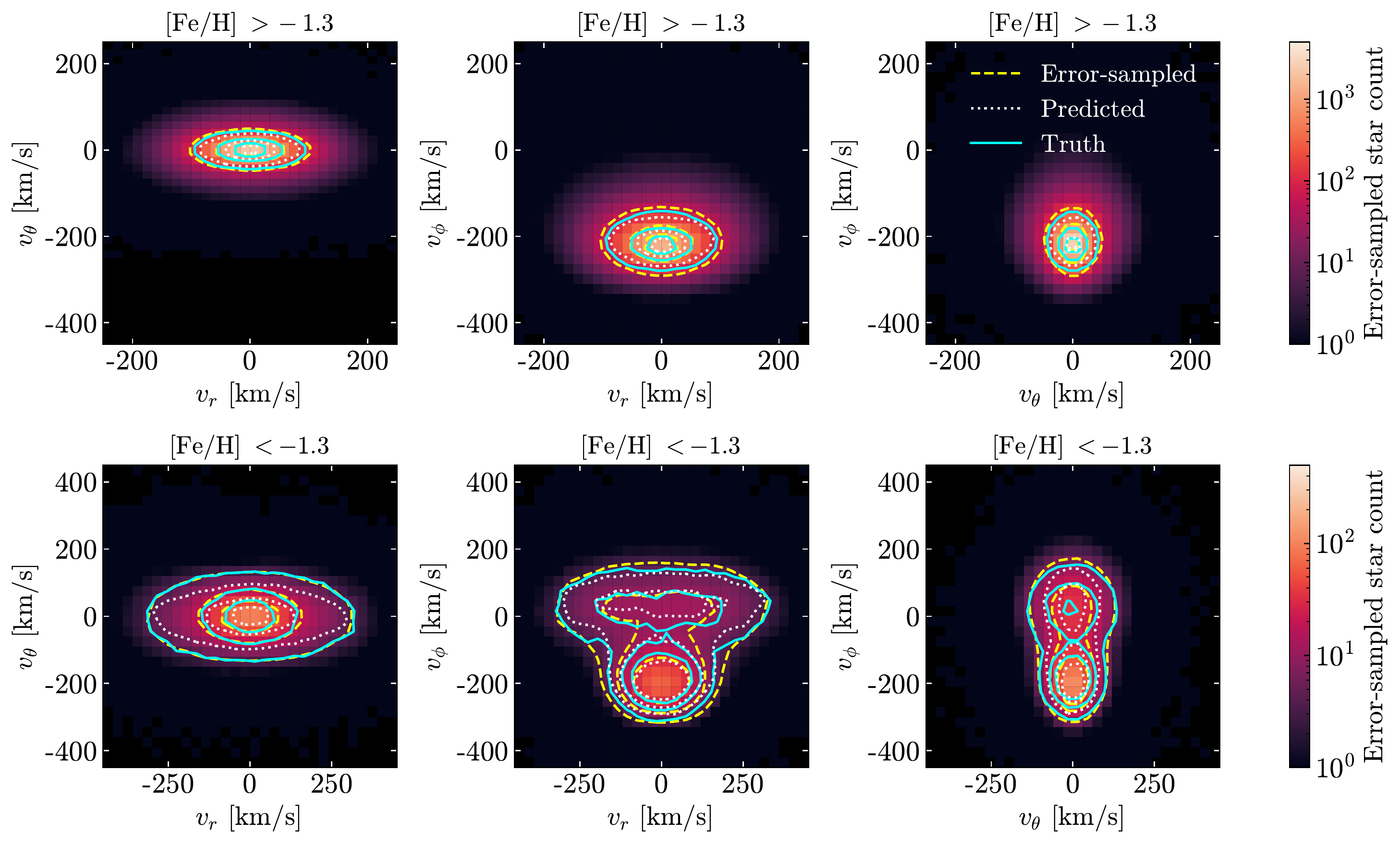}
\caption{Two-dimensional distributions are shown for the Galactocentric velocity components. The background histogram shows the network-predicted kinematic distributions of stars in the test set, sampled over the network’s uncertainty prediction, to obtain the error-sampled distribution.  The top row shows the stars with high-metallicity ([Fe/H] $> -1.3$), and the bottom row shows the stars with low-metallicity ([Fe/H] $< -1.3$) in which the stellar halo and substructure is more prominent. The contours indicate the location of $30\%$, $60\%$, and $90\%$ containment intervals for the true (solid blue), predicted (dotted white) and error-sampled (dashed yellow) distributions. We emphasize that the correlations between metallicity and velocity are being captured by the network, even though the network has no direct access to the metallicity information.}
\label{fig:2dhist}
\end{figure*}

\section{Results} \label{sec:results}

\begin{figure*}[!t]
\centering
\includegraphics[width=\textwidth]{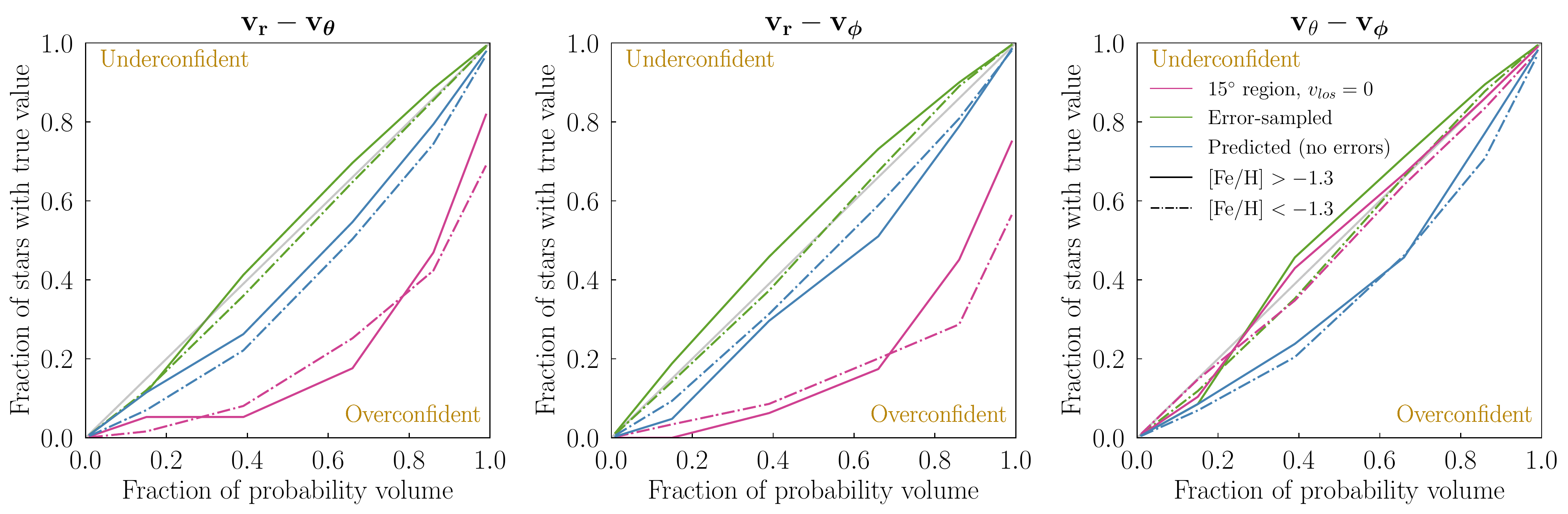}
\caption{The quantile-quantile plots on the neural network posteriors 
in Galactocentric velocity components, showing the fraction of stars with true values that fall into containment intervals defined by the contours of the error-sampled and predicted distributions. The solid green(blue) line and the dashed-dotted green(blue) line correspond to the error-sampled(predicted) contours in the metal-rich ([Fe/H] $> -1.3$) and metal-poor ([Fe/H] $< -1.3$) panels in Fig.~\ref{fig:2dhist}, respectively. In general, the network is well-calibrated for the error-sampled distributions (i.e., the green lines track the diagonal gray line in each panel). For comparison, we compare the network results with those of setting $v_{\text{los}}=0$ for stars that lie in the $15^{\circ}$ region towards and away from the Galactic Center (pink lines)~\citep{koppelman_characterization_2019}.  Restricting to this region of interest produces a stellar sample for which we are relatively sure of $v_{\theta}$ and $v_{\phi}$, as reflected in the calibration in the $v_{\theta} - v_{\phi}$ (third panel). However, the neural network yields a much better calibration in the $v_{r} - v_{\phi}$ and $v_{r} - v_{\theta}$ planes.}
\label{fig:roc}
\end{figure*}

Next, we present the results of applying the neural network to our mock catalog.  A powerful outcome of this procedure is the ability to construct an error-sampled line-of-sight velocity distribution using the network's predicted uncertainty.  We now want to quantify how well the resulting velocity distributions, as well as their correlations with each other, reproduce the truth.  With the network's prediction, we can perform a coordinate transformation on all three velocities into the Galactocentric frame ($v_r$, $v_\theta$, $v_\phi$) and properly assign uncertainties using the procedure in  Appendix~\ref{sec:appendix_velocity}. This makes the presence of the Enceladus-like substructure more readily apparent.

The top row in Fig.~\ref{fig:1dhist} compares the truth, predicted, and error-sampled predicted distributions for $v_{r}, v_{\theta} ,v_{\phi}$.  The error-sampled predicted distribution is determined by the average of 50 Monte Carlo~(MC) trials, which are computed by Gaussian sampling a line-of-sight velocity per star using the network's prediction for its value and uncertainty.  The sampled line-of-sight velocity along with the five known components are subsequently transformed into Galactocentric spherical coordinates. As is apparent in Fig.~\ref{fig:1dhist}, the error-sampled predicted distribution is an excellent approximation of the true distribution.

In the bottom row of Fig.~\ref{fig:1dhist}, the predicted uncertainty on the Galactocentric velocities is shown as a function of the predicted Galactocentric velocity for all stars in the test set, as well as the subset of stars that comprise the substructure.  For each velocity bin, the box denotes the 50\% containment about the median uncertainty, while the whiskers denote the 5\% and 95\% containment.  The predicted uncertainty for a given star is correlated with its spatial location and proper motion (Fig.~\ref{fig:corner_plot}).  In general, we find that the velocity uncertainties for all stars (pink) are lowest for values of $v_r, v_\theta, v_\phi$, where there is a comparatively small contribution of substructure stars (green).  This is also where the velocity uncertainties on the substructure stars are largest, as expected given that the network is forced to distinguish them from the disk stars.  Indeed, we see that that the network erroneously predicts a small tail of substructure stars near $v_\phi^{\rm pred} \sim -200$~km/s.
                          
The spread in the predicted velocity uncertainties in the Galactocentric frame can be substantial.  Depending on the application, one may wish to restrict to a subset of the data with predicted uncertainties in a particular range.  We find that 25\%(13\%) of the 10-million star test set has $\sigma_r^\text{pred} \lesssim$ 10(5)~km/s, while 13\%(6\%) of the $\sim$ 35,000 substructure stars in the test set have $\sigma_r^\text{pred} \lesssim$ 10(5)~km/s. Additionally, we find that 55\%(33\%) of the entire test set and 9\%(4.5\%) of the substructure stars in the test set have $\sigma_\theta^\text{pred} \lesssim$ 10(5)~km/s, and 20\%(10\%) of the entire test set and 10.5\%(5\%) of the substructure stars in the test set have $\sigma_\phi^\text{pred} \lesssim$ 10(5)~km/s.  We have verified that restricting the sample of stars to those with the lowest predicted uncertainties does not bias the kinematic distributions (Figs.~\ref{fig:1D_sigmacut} and~\ref{fig:2D_sigmacut}).

 Figure~\ref{fig:2dhist} explores the extent to which the network captures the correlations among the velocity components.  The top panels show the high-metallicity stars with $\text{[Fe/H]} > -1.3$ (9,944,046 stars), and the bottom panels show the low-metallicity stars with $\text{[Fe/H]} < -1.3$ (55,942 stars).  Dividing the dataset into high- and low-metallicity stars allows us to evaluate the performance of the network on disk and halo/substructure stars, respectively.  To compute the density histograms in Fig.~\ref{fig:2dhist}, we take the mean of 10(500) MC samples drawn from the high-metallicity(low-metallicity) predicted distribution; these are calculated the same way as the error-sampled distributions shown in Fig.~\ref{fig:1dhist}. 

The solid blue(dashed yellow) contours in Fig.~\ref{fig:2dhist} indicate the location of $30\%$, $60\%$, and $90\%$ containment intervals for the true(error-sampled) distributions.  For comparison, the dashed white contours correspond to the predicted distributions without sampling over the network's error prediction; while they do a reasonable job at capturing the 2D velocity correlations, they do not exactly reproduce the truth distributions.  In contrast, the error-sampled distributions nicely track the truth distributions for both the metal-poor and metal-rich samples. This is most striking for the metal-poor sample, where the velocity contours have a non-trivial shape due to contributions from the stellar disk, halo, and substructure.  For example, the network captures the extended $v_r$ distribution at the 60\% containment interval in the low-metallicity $v_{r} - v_{\phi}$ plane, which is a characteristic of the Enceladus-like stars.  It is important to highlight that while Fig.~\ref{fig:2dhist} is separated into metal-rich and metal-poor categories, this is done after training; the network has no direct  access to this information, and yet is able to predict the differences in line-of-sight distributions for these categories.

Figure~\ref{fig:roc} summarizes the network's prediction accuracy in the 3D velocity phase space.  Each panel is a quantile-quantile plot that shows the fraction of stars (according to their true stellar velocities) which fall into containment intervals defined by the contours of the error-sampled and predicted distributions, examples of which are delineated in Fig.~\ref{fig:2dhist}, for different pairs of velocity coordinates. Below the gray line, the network is considered overconfident because a given containment region in the predicted distribution is too small compared to the truth expectation (i.e., the predicted $x\%$ containment region contains $<x\%$ of the true stars).  Above the gray line, the opposite is true and the network is considered underconfident. 

The solid and dash-dotted lines in Fig.~\ref{fig:roc} correspond to the [Fe/H] $> -1.3$ and [Fe/H] $< -1.3$ panels in Fig.~\ref{fig:2dhist}, respectively.  The green lines show the results for the error-sampled distributions.  We see that both the metal-poor and metal-rich distributions are well-calibrated in the $v_{r} - v_{\phi}$,  $v_{r} - v_{\theta}$, and  $v_{\theta} - v_{\phi}$ planes.  For comparison, the blue lines show the same results, but without taking into account the predicted errors.  The network is more overconfident in this case, especially in the $v_{\theta} - v_{\phi}$ plane.  These results reflect the qualitative features of Fig.~\ref{fig:2dhist}.

Without machine learning techniques, reconstructing the full 6D Galactocentric phase space of stars with only 5D information can only be achieved under certain restrictive conditions. For example,~\cite{koppelman_characterization_2019} proposed setting $v_\text{los} = 0$ for stars that lie in the region defined by $\sqrt{\ell^{2} + b^{2}} \leq 15^{\circ}$ or $\sqrt{(\ell - 180^{\circ})^{2} + b^{2}} \leq 15^{\circ}$.  In this region, $v_\theta$ and $v_\phi$ roughly correspond to the proper motion of stars, and can therefore be reconstructed reasonably accurately, with limited dependence on $v_\text{los}$.  Restricting to this region, we have 508,798 metal-rich stars and 1,504  metal-poor stars.  We then compare the predicted velocity distributions, obtained by simply setting $v_\text{los} =0$~km/s, to the truth distribution; however, it is important to note that there is no principled way to assign errors to $v_\text{los}$ with this prescription.  As seen in Fig.~\ref{fig:roc}, the results are well-calibrated and even outperform the network in the $v_{\theta} - v_{\phi}$ plane, where this method is designed to excel. We can see however that the same method is overconfident in the $v_{r} - v_{\phi}$ plane and $v_{r} - v_{\theta}$ plane for both the metal-rich and metal-poor subsets.  Our neural network approach therefore has the benefit of being applicable beyond a narrow spatial region, and can additionally predict uncertainties.  

\section{Conclusions} \label{sec:conlude}

In this Letter, we demonstrated that a neural network can successfully predict a star's line-of-sight velocity and associated uncertainty from 5D astrometric inputs, after being trained on a subset of data with complete 6D phase-space information.  We trained, tested, and validated the network on mock data that contained a disk, stellar halo, and Enceladus-like substructure.  The error-sampled network prediction successfully reproduced the individual velocity distributions, as well as their correlations.  The final results also captured the expected metallicity dependence of the velocity distributions, even though metallicity was not provided as an input to the network.  The network successfully reconstructed the velocities of the substructure stars, even though they comprised only $\sim 0.5\%$ of the training set.  This result demonstrates that the network is learning more than just the bulk motion of the stars.

A critical feature of the network design is its ability to provide an uncertainty on its velocity prediction. 
The predicted velocity distributions more reliably reproduced the true distributions when properly sampled over these errors.  The mean network uncertainty on the predicted line-of-sight velocity was $\sigma_{\rm los}^{\rm pred} = 38$~km/s, with about 8\%(1\%) of all stars in the test set having $\sigma_{\rm los}^{\rm pred} \lesssim 30(20)$~km/s.  We stress that these uncertainties should not be directly compared to the \emph{measured} errors on \gaia's line-of-sight velocities.  In particular, $\sigma_{\rm los}^{\rm pred}$ is the network's \emph{predicted} uncertainty on its projected value of line-of-sight velocity---for instance, a correct network prediction for $v_{\rm los}$ can still be associated with a large uncertainty.  However, we do find that stars with smaller $\sigma_{\rm los}^{\rm pred}$ are typically associated with more accurate velocity predictions.  Furthermore, restricting to a subset of stars with these smaller uncertainties does not bias the overall velocity distributions.  

The simple idea of regressing missing kinematic information can be accomplished by a variety of architecture choices.  For example, Bayesian neural networks~\citep{Mackay, Bhat:2005hq, neal2012bayesian, gal2016dropout, 2016JInst..11P7006B, doi:10.1080/01621459.2017.1285773, Bollweg:2019skg, wagner-carena_hierarchical_2020,charnock2020bayesian} may provide an alternative method for incorporating uncertainties on the network output.  In this case, the network would be rerun many times over the same inputs, while the weights float within some prior distribution.  Since these approaches require a more sophisticated neural network architecture, we leave a detailed comparison for future work.

We optimized the neural network architecture by applying it to a mock \gaia catalog. In practice, the network will be trained on the subset of \gaia data with complete 6D phase space, eliminating any systematic uncertainty on the network output associated with using simulated data. Our training set will expand with \gaia DR3, which will provide line-of-sight velocities for an additional $\sim30$~million stars~\citep{gaiacollaboration2020gaia}. The success of our machine learning approach motivates further studies on other potential applications, such as recovering individual stellar streams and/or multiple substructures that overlap in phase space. 

Thus far, we have trained and tested the network on stars concentrated within 5~kpc of the Sun, but it would also be beneficial to adapt the method presented here to apply to stars farther out in the halo.  This extension would be particularly relevant for  mapping the potential of the Milky Way halo and identifying dwarf galaxies and other substructures, as just two important examples.  For current \gaia data releases, the primary challenge is the lack of a sufficient training sample, limited by the number of stars at these distances with complete 6D information.  One possibility is to train the network on a subset of \gaia stars whose line-of-sight velocities come from cross-matches to other spectroscopic surveys.  Whether this approach will succeed depends on both the number of cross-matched stars and whether the selection function of the spectroscopic surveys complicates the network inference.  If successful,  this machine learning approach can potentially be adapted to the upcoming Rubin Observatory. 

This Letter shows that there is often incredible information waiting to be extracted from high dimensional correlations, which can be used to maximize the utility of real world data.

\section*{Acknowledgements} \label{sec:acknowledge}
The authors acknowledge  H.~Koppelman, P.~Melchior, S.~Mishra-Sharma, L.~Necib, O.~Slone, and N.~Weiner for fruitful conversations.  ML gratefully acknowledges financial support from the Schmidt DataX Fund at Princeton University made possible through a major gift from the Schmidt Futures Foundation.  BO was supported in part by the U.S. Department of Energy~(DOE) under contract DE-SC0013607 and DE-SC0020223. LJC was supported in part by a Paul \& Daisy Soros Fellowship and the NSF GRFP Award Number DGE-1656466. TC is supported by the DOE under Award Number  DE-SC0011640. HL and ML are supported by the DOE under Award Number DE-SC0007968.  The work presented in this paper was performed on computational
resources managed and supported by Princeton Research
Computing. 
This work is supported by the National Science Foundation under Cooperative Agreement PHY-2019786 (The NSF AI Institute for Artificial Intelligence and Fundamental Interactions, http://iaifi.org/).
This research made use of the \texttt{astropy}~\citep{Robitaille:2013mpa}, \texttt{corner}~\citep{corner}, \texttt{h5py}~\citep{collette_python_hdf5_2014},  
\texttt{IPython}~\citep{PER-GRA:2007}, 
\texttt{Jupyter}~\citep{Kluyver2016JupyterN}, \texttt{matplotlib}~\citep{Hunter:2007}, 
\texttt{NumPy}~\citep{numpy:2011}, \texttt{pandas}~\citep{mckinney-proc-scipy-2010}, and 
\texttt{SciPy}~\citep{Jones:2001ab} software packages.

\clearpage
\appendix

\setcounter{figure}{0} \renewcommand{\thefigure}{A\arabic{figure}} \renewcommand{\theHfigure}{A\arabic{figure}}
\setcounter{table}{0} \renewcommand{\thetable}{A\arabic{table}} \renewcommand{\theHtable}{A\arabic{table}}
\setcounter{equation}{0} \renewcommand{\theequation}{A\arabic{equation}} \renewcommand{\theHequation}{A\arabic{equation}}

In Appendix~\ref{sec:appendix_network}, we elaborate on the details of our network architecture, as well as its optimization. In Appendix~\ref{sec:appendix_velocity}, we explain how to transform the line-of-sight velocity and uncertainty provided by the network to Galactocentric velocities and associated uncertainties. Finally, in Appendix~\ref{sec:appendix_figures}, we provide several supplemental figures that are referenced in the main text. 

\section{Network information}\label{sec:appendix_network}
\subsection{Tested Network Architectures}\label{tested_architectures}

\begin{figure}[t]
    \centering
    \includegraphics[width=0.5\textwidth]{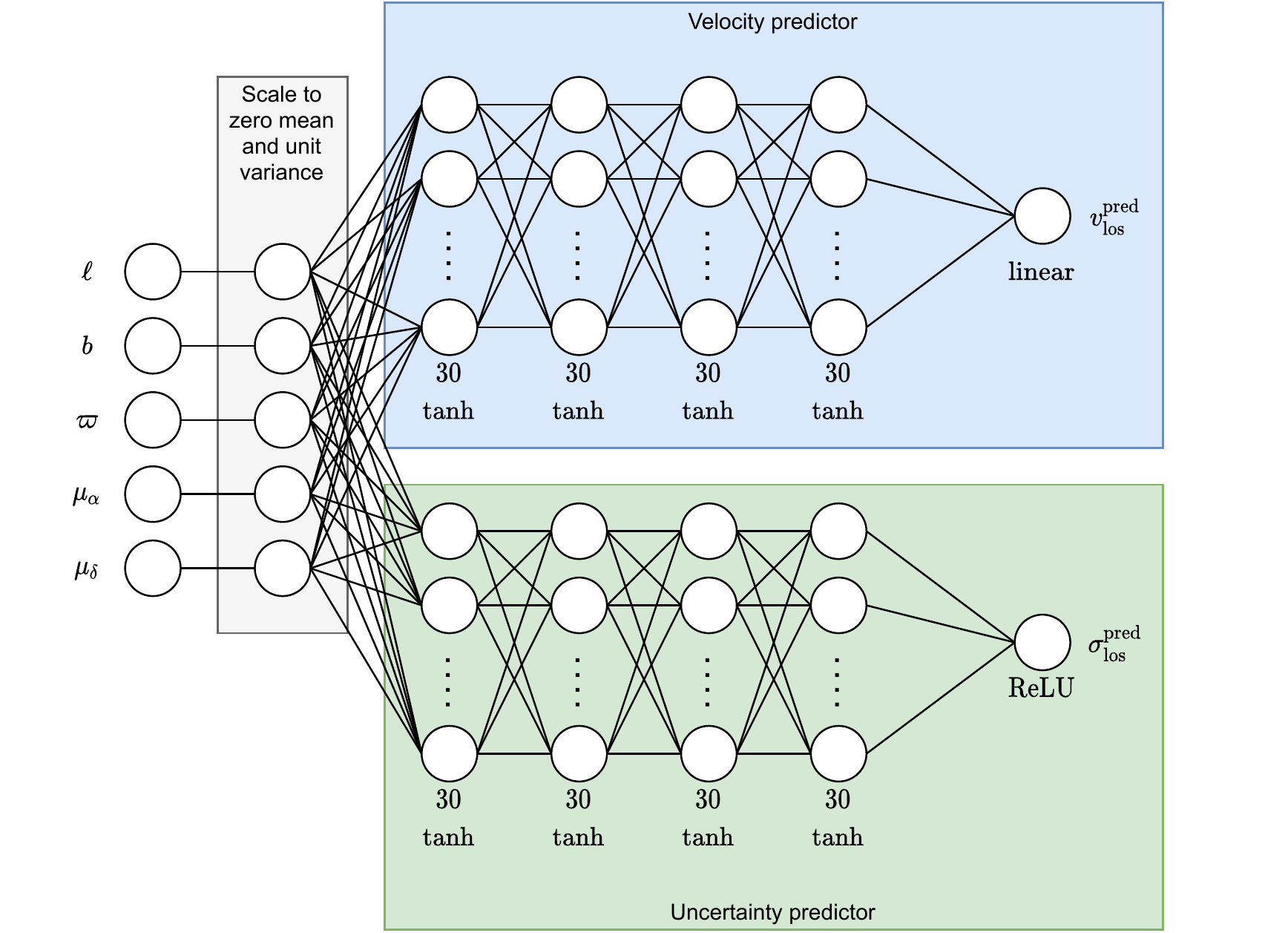}
    \caption{Diagram of the neural network used in this work. }
    \label{fig:networks}
\end{figure}

In this appendix, we elaborate on the procedure used to select the network architecture described in Sec.~\ref{sec:network}.   The final configuration, illustrated in Fig.~\ref{fig:networks}, was chosen after extensive tests that varied the number of hidden layers, the number of nodes per layer, the activation function associated with each layer, the sample weights, and the input parameters. 

We tested network architectures under  combinations of the following specifications: 1--5 hidden layers, 30--2000 nodes per layer, ReLU/ELU/Tanh activation functions on the interior layers, and different combinations of input parameters. Deeper (more layers) and wider (more nodes per layer) networks do not yield a significantly more accurate regression, and conversely, smaller and narrower networks both perform a less accurate regression. Neither the ReLU activation function nor the ELU activation function predict the velocity distribution as accurately as the Tanh activation function.

In order to evenly sample the tails of the $v_\text{los}$ distribution in our dataset, we implement sample weighting to de-emphasize the contribution of stars with $v_\text{los}\sim 0\,\mathrm{km/s}$. The simplest way to effectively ``flatten" the peak of the $v_\text{los}$ distribution is to take the $v_\text{los}$ distribution itself to be the reciprocal of the weighting function. In practice, we histogram the $v_\text{los}$ values in each dataset, then interpolate the histogram to extract the probability $p_{i}$ corresponding to a given star's value of $v_{\text{los},i}$ and take $w_{i}=1/p_{i}$ to be the weight for that star. We refer to this as the ``linear weights" prescription, to contrast with the ``logarithmic weights" prescription described below.

We can also implement a weighting scheme that is a less-steep function of $v_{\text{los},i}$ than $1/p_{i}$. To do so, we define the ``logarithmic weights" as $w_{i}=\log(1/p_{i})+ w_{0}$, where $w_{0}$ is chosen so that $w_{i}$ is positive for all stars. In practice, we find that linear weights in $v_\text{los}$ overly de-emphasize the central $v_\text{los}$ values, while logarithmic weights alleviate this issue. Given the dependence of $v_\text{los}$ on Galactic longitude,~$\ell$, we also implement linear and logarithmic weights on the joint distribution of $v_\text{los}$ and $\ell$.   Of the four weighting schemes tested, the joint $v_\text{los}$--$\ell$ logarithmic weights method is the most effective at recovering the truth distributions.  We  thus choose this to be our fiducial method for sample weighting.

We also tested a network that included the information from each star's ten nearest neighbors during training. 

Because this additional information does not improve the regression, at least for the dataset considered here, we do not adopt it for the fiducial method.  Lastly, we compared our combined network structure, described in Sec.~\ref{sec:network}, to a single network with two outputs, and found that the latter does not not learn or train well.  

\subsection{Network Diagnostics}\label{network_diagnostics}
The network architectures described above in Sec.~\ref{tested_architectures} were evaluated using the Kullback-Leibler (KL) divergence, as well as the $R^{2}$ and $\chi^{2}$ test statistics. These statistics are defined as follows:
\begin{equation}
    \mathcal{D}_{KL}(\textbf{p}||\textbf{q}) = \sum_{s}p_{s}\, \text{log}_{2}\left(\frac{p_{s}}{q_{s}}\right) \, , \quad
    R^{2} = 1 - \sum_{i=1}^{N} \left(\frac{v_{\text{los},i}-v^{\text{pred}}_{\text{los},i}}{v_{\text{los},i}  - \Bar{v}_{\text{los}}}\right)^2 
    \,, \text{ and }\quad     
    \chi^{2} = \frac{1}{N-1}\sum_{i=1}^{N} \left(\frac{v_{\text{los},i}-v^{\text{pred}}_{\text{los},i}}{ \sigma^{\text{pred}}_{\text{los},i}} \right)^2 \, ,
    \label{eq:KL}
\end{equation}
where $v_{\text{los},i}$, $v^{\text{pred}}_{\text{los},i}$, and $\sigma^{\text{pred}}_{\text{los},i}$ are defined as in Eq.~(\ref{eq:lossfunc}), and $\Bar{v}_{\text{los}}$ is the mean of the true line-of-sight velocity  distribution.  Also, $\textbf{p}$, $\textbf{q}$ are the values of the kernel density estimation of the true and predicted distributions, respectively, evaluated at each histogram bin value $s$ $\in [-350,350]$ km/s in Fig.~\ref{fig:1dhist}. 

The KL divergence is a measure of how one probability distribution differs from a second, reference probability distribution~\citep{kullback_1951}. A small value of the KL divergence $\mathcal{D}_{KL}$ indicates very good agreement between the true and predicted distributions. In other words, if the probability for an event from \textbf{p} is large, but the probability for the same event in \textbf{q} is small, there will be a large contribution to the statistic. The $R^{2}$ test statistic provides a measure of how well the truth values are replicated by the predicted values, without including the predicted uncertainty measure, $\sigma^{\text{pred}}_{\text{los}}$. The $R^{2}$ statistic generally ranges from 0 to 1, but can be negative if the fit is very poor.  $R^{2} = 0$ is obtained if the predicted velocity is always the mean velocity, while $R^{2} = 1$ indicates a completely accurate prediction for all stars. The $\chi^{2}$ test statistic is a test of the uncertainty prediction, $\sigma_{\text{los}}^\text{pred}$. In general, if $\chi^{2} \gg 1$, then the predicted uncertainty is too small. Conversely, if $\chi^{2} \ll 1$, then the predicted uncertainty is too large. We note that each predicted uncertainty is approximately  Gaussian. That is, the quantity $\left(v_{\text{los}, i} - v_{\text{los}, i}^\text{pred} \right)/\sigma_{\text{los}, i}^\text{pred}$ for the $i^\text{th}$ star is approximately Gaussian-distributed over the full sample. 

We trained and tested the different network architectures on a smaller dataset than the one ultimately used in this work and analyzed these architectures using the test statistics in Eq.~\eqref{eq:KL}.  We then calculated the test statistics for each case, evaluating how sensitive the results are to various cuts that span $\sigma^{\text{pred}}_{\text{los}} < 30$~km/s to $\sigma^{\text{pred}}_{\text{los}} < 150$~km/s.  The results are provided in Table~\ref{table:1}; all networks included in this table utilize the logarithmic sample weights on the joint distribution of $v_{\text{los}}$ and $\ell$.

\begin{table*}[b]
\centering
\footnotesize
\renewcommand{\arraystretch}{1.5}
\noindent
\makebox[\textwidth][c]{\resizebox{\linewidth}{!}{ }}
\begin{tabular}{ C{3 cm} | C{2 cm} |  C{2 cm}  C{2 cm}  C{2 cm}}
 \text{Input Variables} & Dropout & KL & $\chi^{2}$  & $R^{2}$ \\
  \Xhline{3\arrayrulewidth}
 $\ell, b, \varpi, \mu_{\alpha}, \mu_{\delta}, x, y, z$ & No & 0.008--0.010 & 0.699--0.749 & 0.720--0.746 \\ 
  \hline
 $\ell, b, \varpi, \mu_{\alpha}, \mu_{\delta}$ & No & 0.008--0.010 & 0.685--0.739 & 0.704--0.721\\  
  \hline
 $x,y,z$ & No & 0.009--0.035 & 0.578--0.779 &  0.681--0.738 \\
  \hline
 $\ell, b, \varpi, \mu_{\alpha}, \mu_{\delta}, x, y, z$ & Yes & 0.002--0.986 & 0.245--4.214 & 0.713--0.857\\ 
  \hline
 $\ell, b, \varpi, \mu_{\alpha}, \mu_{\delta}$ & Yes & 0.017--0.075 & 0.348--0.789 & 0.712--0.757 \\  
  \hline
 $x,y,z$ & Yes & 0.008--0.096 & 0.314--3.040 & 0.651--0.818\\
\end{tabular}

\caption{This table summarizes a few network architectures studied in this work,  and the range of test statistics obtained for each case when restricting the dataset by making various cuts that span $\sigma^{\text{pred}}_{\text{los}} < 30$~km/s to $\sigma^{\text{pred}}_{\text{los}} < 150$~km/s.  The test statistics are defined in Eq.~\eqref{eq:KL}. Our fiducial model is the second row of the table.}
\label{table:1}
\end{table*}

Dropout is a technique in which randomly selected nodes are ignored when the network is trained. When not using dropout and including the proper motions in the input, the results are not sensitive to cuts on the predicted uncertainty (i.e., the range of the test statistic is small). This is very desirable because the goodness-of-fit does not depend on the particular selection of stars. However, the networks with dropout have a wide range in the diagnostic metrics as cuts on $\sigma_\text{los}^\text{pred}$ are applied. Using such networks would thus require extra calibration to understand predictions when departing from the nominal set of stars.

In the case without dropout, the test statistics and ranges for the $(\ell, b, \varpi, \mu_{\alpha}, \mu_{\delta}, x, y, z)$ and $(\ell, b, \varpi, \mu_{\alpha}, \mu_{\delta})$ configurations are very similar, suggesting that including the redundant $(x, y, z)$ inputs does not aid the network.  

In general, the $(x,y,z)$ network does worse overall, especially when using dropout. This provides a strong indication that the network is learning more than just the bulk stellar motion, because it uses more than just the location in the Galaxy.

When dropout is included, the network does better on the initial predictions (i.e., the $R^{2}$ and KL divergence have better scores than without dropout at one end of the range), but the spread of the test statistics is still very large. Also, when looking at $\chi^{2}$ for these models, we see that the estimate of the uncertainty is less accurate, and the spread of these values is much greater.

After running numerous tests on network architectures, activation functions, input data, and weighting schemes, we ultimately selected a network with 4 hidden layers with 30 nodes each and without dropout. We only include the 5D coordinates $\ell, b, \varpi, \mu_{\alpha}, \mu_{\delta}$ as input because the extra variables do not significantly help with the regression.

\begin{figure}[t]
    \centering
    \includegraphics[width=\textwidth]{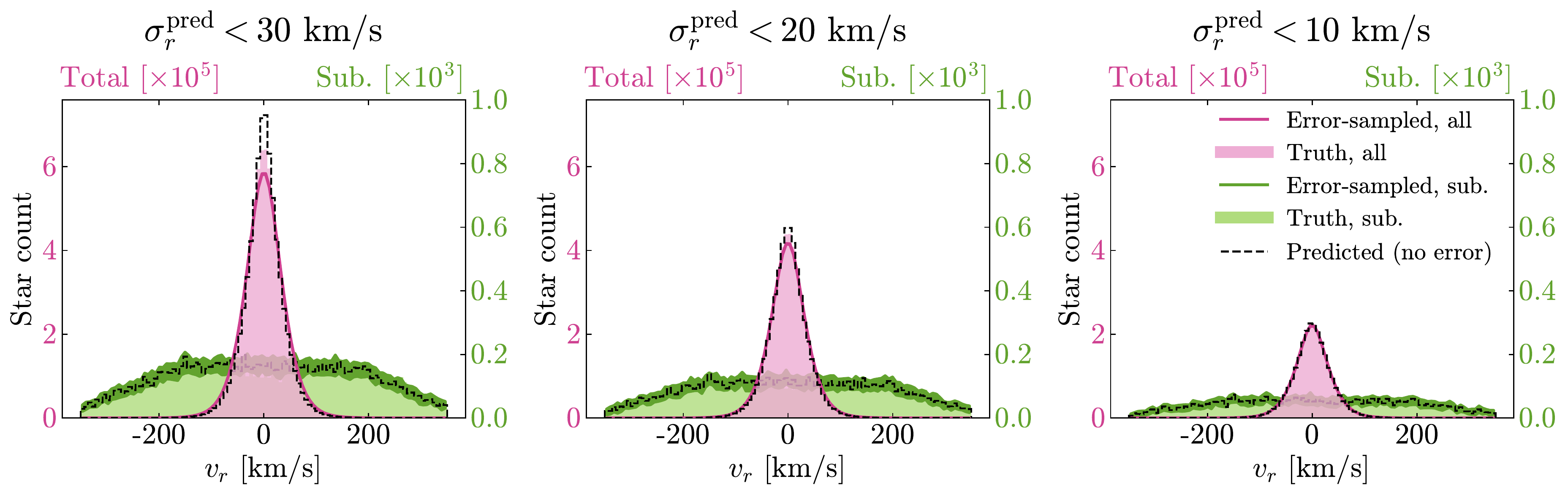}
    \caption{The lines depicted in this figure are the same as those in Fig. \ref{fig:1dhist}. However, this figure specifically focuses on the Galactocentric radial velocity $v_{r}$, to show that the distribution is unbiased as more restrictive cuts on $\sigma_{r}^{\text{pred}}$ are made.}
    \label{fig:1D_sigmacut}
\end{figure}

As described in the main text, Fig.~\ref{fig:1dhist} highlights the significance of including predicted uncertainty by comparing the truth, predicted, and error-sampled distributions of $v_{r}$, the Galactocentric radial velocity. The test statistics described above in Sec.~\ref{network_diagnostics} were calculated for the $v_{r}$ histograms containing all test stars.  When $\sigma^{\text{pred}}_{r}$ is included, the KL divergence is 0.009, compared to 0.016 when $\sigma^{\text{pred}}_{r}$ is not included.  Figure~\ref{fig:1D_sigmacut} shows the same Galactocentric radial velocity histograms as in Fig.~\ref{fig:1dhist}, but selecting stars with $\sigma_r^\text{pred} < 30, 20, 10$~km/s.  The KL divergence between the truth and error-sampled distributions decreases from $0.005$ when $\sigma^{\text{pred}}_{r} < 30$~km/s, to $0.002$ when $\sigma^{\text{pred}}_{r} < 20$~km/s, to $< 0.001$ when $\sigma^{\text{pred}}_{r} < 10$~km/s. Making increasingly stronger cuts on uncertainties in a certain velocity component can be used to obtain increasingly more accurate distributions in that component. It is also possible to make an uncertainty cut on two coordinates, for example $v_{r}$ and $v_{\phi}$, in pursuit of a high-purity sample of Enceladus-like stars. 

The predicted $v_r$ values are also evaluated using the $\chi^{2}$ and $R^{2}$ test statistics. The value of $\chi^{2}$ for the test set is 0.68 and remains stable when cuts on $\sigma^{\text{pred}}_{r}$ are made, from $0.679$ when $\sigma^{\text{pred}}_{r} < 30$~km/s, to $0.663$ and $0.641$ when $\sigma^{\text{pred}}_{r} < 20$~and $10$~km/s, respectively.  The value of $R^{2}$ is 0.72. As cuts on $\sigma^{\text{pred}}_{r}$ are made, the value of $R^{2}$ approaches 1.0, from $0.862$ when $\sigma^{\text{pred}}_{r} < 30$~km/s, to $0.937$ and $0.985$ when $\sigma^{\text{pred}}_{r} < 20$ and $10$~km/s, respectively.  This again highlights the usefulness of the network's uncertainty prediction in attaining accurate velocity predictions. The correlations for the set of stars with $\sigma^{\text{pred}}_{r} < 10$~km/s are shown in Fig.~\ref{fig:2D_sigmacut}, in which the truth, predicted and error-sampled distributions align very well in $v_{r}$. As expected, we see that the truth and error-sampled distributions across all velocity pairs align better overall. 

\clearpage

\setcounter{figure}{0} \renewcommand{\thefigure}{B\arabic{figure}} \renewcommand{\theHfigure}{B\arabic{figure}}
\setcounter{table}{0} \renewcommand{\thetable}{B\arabic{table}} \renewcommand{\theHtable}{B\arabic{table}}
\setcounter{equation}{0} \renewcommand{\theequation}{B\arabic{equation}} \renewcommand{\theHequation}{B\arabic{equation}}

\section{Transforming Velocities and Uncertainties}\label{sec:appendix_velocity}
In this appendix, we derive the affine transformation that maps proper motions and line-of-sight velocity to Galactocentric velocities. As a corollary, we also obtain the transformation mapping velocity uncertainties between these respective coordinate systems.

Consider a star with equatorial coordinate velocities, given by $[v_\text{los}, \dot{\alpha} d \cos \delta, \dot{\delta} d]^\intercal$, where $\alpha$ and $\delta$ are right ascension and declination, respectively, $v_\text{los}$ is the line-of-sight velocity and $d$ is the distance of the star from the Sun. The dot represents a derivative with respect to time. With respect to previously defined variables in the text, we have $\mu_\alpha = \dot{\alpha} d \cos \delta$, $\mu_\delta = \dot{\delta} d$, and $d = 1/\varpi$. Following~\cite{2011PhDT........86B}, equatorial coordinate velocities can be transformed into Galactic coordinates velocities via
\begin{alignat}{1}
    \begin{bmatrix}
        v_\text{los} \\
        \dot{\ell} d \cos b \\
        \dot{b} d
    \end{bmatrix} = \begin{pmatrix}
        1 & 0 & 0 \\
        0 & \cos \psi  & \sin \psi \\
        0 & -\sin \psi & \cos \psi 
    \end{pmatrix} \begin{bmatrix}
        v_\text{los} \\
        \dot{\alpha} d \cos \delta \\
        \dot{\delta} d
    \end{bmatrix} \equiv \mathsf{P} \begin{bmatrix}
        v_\text{los} \\
        \dot{\alpha} d \cos \delta \\
        \dot{\delta} d
    \end{bmatrix} \,,
    \label{eq:proper_motion_to_galactic_velocities}
\end{alignat}
where $\psi \simeq 78.3433^\circ$ is the Galactic parallactic angle. 

Next, we want to convert velocities in Galactic coordinates into Galactocentric coordinates. To do this, we begin by writing down the relation between Cartesian coordinates in each coordinate system. Consider a star with Cartesian Galactocentric coordinates $\vec{r}_\text{GC}$; its coordinates $\vec{r}_\text{g}$ in Cartesian Galactic coordinates is given by
 \begin{alignat}{1}
     \vec{r}_\text{g} = \mathsf{M}_\odot^{-1} \left(\vec{r}_\text{GC} - \vec{R}_{\odot,\text{GC}} \right) \,,
     \label{eq:relation_btwn_galactic_and_GC_coords}
 \end{alignat}
 where $\vec{R}_{\odot,\text{GC}} = (x_{\odot,\text{GC}}, 0, z_{\odot,\text{GC}})^\intercal$ is the Cartesian Galactocentric coordinates of the Sun (we adopt the standard convention that the Sun lies in the $xz$-plane in these coordinates, with the Galactic plane lying in the $xy$-plane). The matrix $\mathsf{M}_\odot$ rotates the Galactic coordinate axes such that they become parallel to the Galactocentric coordinate axes, and is written explicitly as
 \begin{alignat}{1}
     \mathsf{M}_\odot^{-1} = \begin{pmatrix}
         \cos \beta & 0 & -\sin \beta \\
         0 & 1 & 0 \\
         \sin \beta & 0 & \cos \beta
     \end{pmatrix} \,,
 \end{alignat}
 where $\beta \equiv \arctan(z_{\odot,\text{GC}} / x_{\odot,\text{GC}})$. Taking the derivative of the coordinates, we find that
 \begin{alignat}{1}
     \dot{\vec{r}}_\text{g} = \mathsf{M}_3 \mathsf{M}_4 \begin{bmatrix}
         v_\text{los} \\ \dot{\ell} d \cos b \\ \dot{b} d
     \end{bmatrix} \,,
 \end{alignat}
with
 \begin{alignat}{1}
     \mathsf{M}_3 = \begin{pmatrix}
         \cos \ell & - \sin \ell & 0 \\
         \sin \ell & \cos \ell & 0 \\
         0 & 0 & 1
     \end{pmatrix} \,, \qquad 
     \mathsf{M}_4 = \begin{pmatrix}
         \cos b & 0 & - \sin b \\
         0 & 1 & 0 \\
         \sin b & 0 & \cos b
     \end{pmatrix} \,.
 \end{alignat}
 Similarly, 
 \begin{alignat}{1}
     \dot{\vec{r}}_\text{GC} = \mathsf{M}_2^{-1} \mathsf{M}_1^{-1} \begin{bmatrix}
         v_r \\ v_\phi \\ v_\theta
     \end{bmatrix} \,,
 \end{alignat}
 with
 \begin{alignat}{1}
     \mathsf{M}_2^{-1} = \begin{pmatrix}
         \cos \phi & -\sin \phi & 0 \\
         \sin \phi & \cos \phi & 0  \\
         0 & 0 & 1
     \end{pmatrix} \,, \qquad
     \mathsf{M}_1^{-1} = \begin{pmatrix}
         \cos \theta & 0 & -\sin \theta \\
         0 & 1 & 0 \\
         \sin \theta & 0 & \cos \theta
     \end{pmatrix} \,.
 \end{alignat}
Finally, taking the derivative of Eq.~\eqref{eq:relation_btwn_galactic_and_GC_coords} and using the expressions derived above, we arrive at the following transformation between Galactic and Galactocentric velocities: 
\begin{alignat}{1}
    \begin{bmatrix}
         v_r \\ v_\phi \\ v_\theta
     \end{bmatrix} = \mathsf{M} \begin{bmatrix}
         v_\text{los} \\ \dot{\ell} d \cos b \\ \dot{b} d
     \end{bmatrix} + \mathsf{M}_1 \mathsf{M}_2 \vec{v}_{\odot,\text{GC}} \,,
\end{alignat}
where we have defined $\mathsf{M} = \mathsf{M}_1 \mathsf{M}_2 \mathsf{M}_\odot \mathsf{M}_3 \mathsf{M}_4$, while $\vec{v}_{\odot,\text{GC}}$ is the Cartesian velocity of the Sun in the Galactocentric frame. Combining this with Eq.~\eqref{eq:proper_motion_to_galactic_velocities}, we obtain the full transformation mapping equatorial coordinate velocities to Galactocentric coordinate velocities:
\begin{alignat}{1}
    \begin{bmatrix}
        v_r \\ v_\phi \\ v_\theta
    \end{bmatrix} = \mathsf{M} \mathsf{P} \begin{bmatrix}
        v_\text{los} \\ \dot{\alpha} d \cos \delta \\ \dot{\delta} d
    \end{bmatrix} + \mathsf{M}_1 \mathsf{M}_2 \vec{v}_{\odot,\text{GC}} \,.
\end{alignat}
This result has been verified by comparison with coordinate transformations implemented by \texttt{Astropy}~\citep{Robitaille:2013mpa}.  

We now turn our attention to the transformation of uncertainties. Assuming Gaussian uncertainties, the differential probability $P(\vec{v}_\text{true}; \vec{v}_\text{m})$ that the true velocity of a star is $\vec{v}_\text{true}(v_\text{los},\mu_\alpha,\mu_\delta)$ given a measured velocity $\vec{v}_\text{m}(v_\text{los},\mu_\alpha,\mu_\delta)$ is proportional to
\begin{alignat}{1}
    P(\vec{v}_\text{true} ; \vec{v}_\text{m}) \propto \exp \left[- \frac{1}{2} \left(\vec{v}_\text{true} - \vec{v}_\text{m}\right)^\intercal \mathsf{S}^{-1} \left(\vec{v}_\text{true} - \vec{v}_\text{m}\right) \right] \,,
\end{alignat}
where $\mathsf{S}(v_\text{los},\mu_\alpha,\mu_\delta)$ is the covariance matrix in the basis of equatorial coordinate velocities. Since the differential probability is invariant under coordinate transformations, we can see immediately that the covariance matrix in the basis of Galactocentric velocities can be obtained by
\begin{alignat}{1}
    \mathsf{S}(v_r, v_\phi, v_\theta) = \mathsf{M} \mathsf{P} \cdot \mathsf{S}(v_\text{los},\mu_\alpha,\mu_\delta) \cdot \mathsf{P}^{-1} \mathsf{M}^{-1} \,.
\end{alignat}

\clearpage
\setcounter{figure}{0} \renewcommand{\thefigure}{C\arabic{figure}} \renewcommand{\theHfigure}{C\arabic{figure}}
\setcounter{table}{0} \renewcommand{\thetable}{C\arabic{table}} \renewcommand{\theHtable}{C\arabic{table}}
\setcounter{equation}{0} \renewcommand{\theequation}{C\arabic{equation}} \renewcommand{\theHequation}{C\arabic{equation}}

\section{Supplementary Figures}\label{sec:appendix_figures}
In this appendix, we provide some figures that supplement the discussion in the main text.  

\begin{figure}[h]
    \centering
    \includegraphics[width=\textwidth]{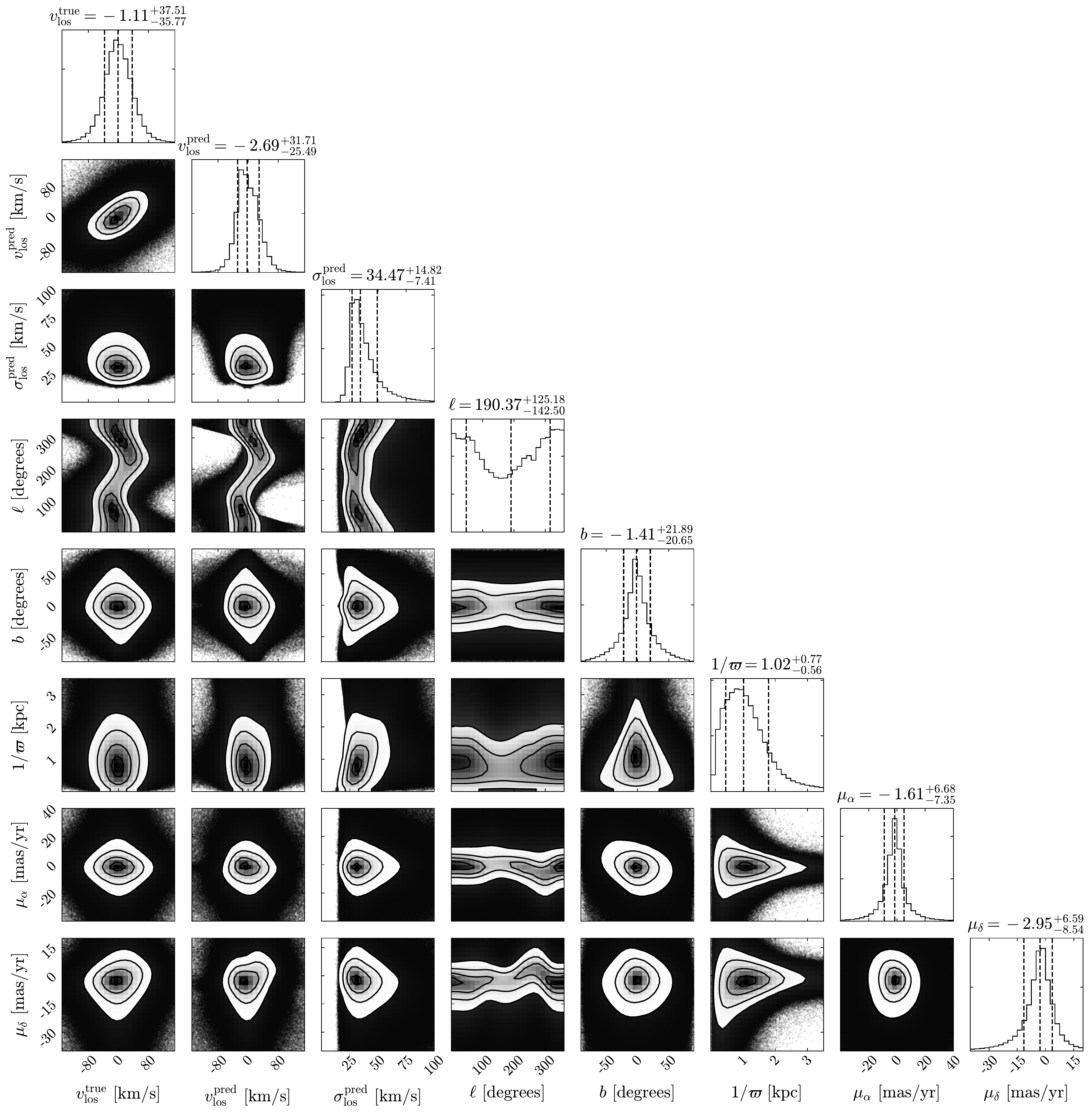}
    \caption{Corner plot for all stars in the test set for the following quantities: true line-of-sight velocity ($v_{\rm los}^{\rm true}$), predicted line-of-sight velocity ($v_{\rm los}^{\rm pred}$), predicted uncertainty on light-of-sight velocity ($\sigma_{\rm los}^{\rm pred}$), Galactic longitude and latitude ($\ell$,$b$), inverse parallax ($1/\varpi$), and proper motions ($\mu_{\alpha,\delta}$).}
    \label{fig:corner_plot}
\end{figure}

\begin{figure}[h]
    \centering
    \includegraphics[width=\textwidth]{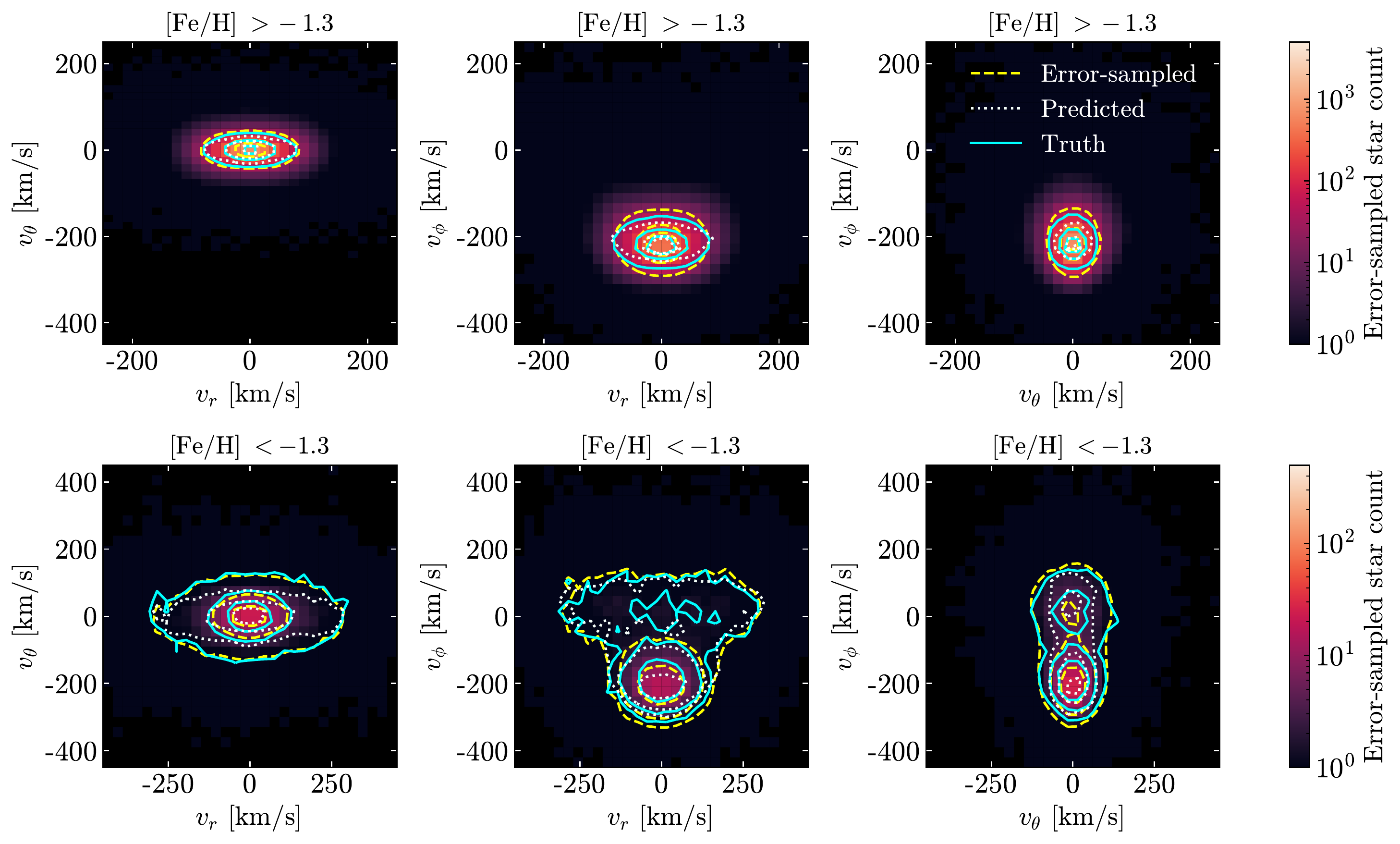}
    \caption{The background histogram and contours depicted in this figure are the same as those in Fig. \ref{fig:2dhist}.  However, this figure shows the resulting error-sampled background histogram and contours for the subset of stars with $\sigma_{r}^{\text{pred}} < 10$~km/s. There are 2,490,330 stars with [Fe/H] $> -1.3$ and 9,316 stars with [Fe/H] $< -1.3$.}
    \label{fig:2D_sigmacut}
\end{figure}

\clearpage
\def\bibsection{} 
\bibliographystyle{aasjournal}
\bibliography{dataX_gaia}

\end{document}